\documentclass[journal]{IEEEtran}
\ifCLASSINFOpdf
 \usepackage[pdftex]{graphicx}
\else
\fi
\usepackage{amsmath}
\usepackage{textcomp}
\usepackage{algorithm}
\usepackage[noend]{algpseudocode}
\usepackage[compact]{titlesec}

\algrenewcommand\algorithmicforall{\textbf{foreach}}
\algrenewcommand\algorithmicindent{.8em}

\usepackage{array,multirow}

\usepackage{array}
\begin{document}
%
\title{Training a quantum annealing based restricted Boltzmann machine on cybersecurity data}
%
%
%

\author{Vivek~Dixit, Raja~Selvarajan,  Tamer~Aldwairi, Yaroslav~Koshka, Mark~A.~Novotny,  Travis~S.~Humble,  Muhammad~A.~Alam, and Sabre~Kais%
\thanks{© 20XX IEEE.  Personal use of this material is permitted.  Permission from IEEE must be obtained for all other uses, in any current or future media, including reprinting/republishing this material for advertising or promotional purposes, creating new collective works, for resale or redistribution to servers or lists, or reuse of any copyrighted component of this work in other works.}%
\thanks{Vivek Dixit, Raja Selvarajan, and Sabre Kais are with the Department of Chemistry, 
Department of Physics and Astronomy, and Purdue Quantum Science and Engineering Institute,
 Purdue University, West Lafayette, IN, 47907, USA}%
\thanks{Muhammad~A.~Alam is with the Department of Electrical and Computer Engineering, 
Purdue University, West Lafayette, IN, 47907, USA }%
\thanks{Travis~S.~Humble is with Quantum Computing Institute, Oak Ridge National Laboratory,
 Oak Ridge, TN, USA}
\thanks{Tamer~Aldwairi is with the Department of Computer and Information Sciences, Temple University, Philadelphia, PA 19122, USA}%
\thanks{Y. Koshka is with the Department of Electrical and Computer Engineering, HPC2 Center for Computational Sciences, Mississippi State University, Mississippi State, MS 39762 USA.
}%
\thanks{M. A. Novotny is with the Department of Physics and Astronomy, HPC2
Center for Computational Sciences, Mississippi State University, Mississippi State,
MS 39762 USA.}%
\thanks{ This manuscript has been authored by UT-Battelle, LLC, under Contract No.~DE-AC0500OR22725 with the U.S.~Department of Energy. The United States Government retains and the publisher, by accepting the article for publication, acknowledges that the United States Government retains a non-exclusive, paid-up, irrevocable, worldwide license to publish or reproduce the published form of this manuscript, or allow others to do so, for the United States Government purposes. The Department of Energy will provide public access to these results of federally sponsored research in accordance with the DOE Public Access Plan.}%
\thanks{Manuscript received XXXX; revised XXXX. 
Corresponding author: Sabre~Kais (email: kais@purdue.edu).}}

%
%

\markboth{IEEE TRANSACTIONS ON EMERGING TOPICS IN COMPUTATIONAL INTELLIGENCE}%
{Shell \MakeLowercase{\textit{et al.}}: Bare Demo of IEEEtran.cls for IEEE Journals}
%



\maketitle

\begin{abstract}
A restricted Boltzmann machine (RBM) is a generative model that could be used in effectively  balancing a cybersecurity dataset  because the synthetic data a RBM generates follows the probability distribution of the training data. RBM training can be performed using contrastive divergence (CD) and quantum annealing (QA). QA-based RBM training is fundamentally different from CD and requires samples from a quantum computer. 
We present a real-world application that uses a quantum computer. Specifically, we train a RBM using QA for cybersecurity applications. The D-Wave 2000Q has been used to implement QA. RBMs are trained on the ISCX data, which is a benchmark dataset for cybersecurity. For comparison, RBMs are also trained using CD. CD is a commonly used method for RBM training. Our analysis of the ISCX data shows that the dataset is imbalanced. We present two different schemes to balance the training dataset before feeding it to a classifier. The first scheme is based on the undersampling of benign instances. The imbalanced training dataset is divided into five sub-datasets that are trained separately. A majority voting is then performed to get the result. Our results show the majority vote increases the classification accuracy up from 90.24\% to 95.68\%, in the case of CD. For the case of QA, the classification accuracy increases from 74.14\% to 80.04\%. In the second scheme, a  RBM is used to generate synthetic data to balance the training dataset.  We show that both QA and CD-trained RBM can be used to generate useful synthetic data. Balanced training data is used to evaluate several classifiers. Among the classifiers investigated, K-Nearest Neighbor (KNN) and Neural Network (NN) perform better than other classifiers. They both show an accuracy of 93\%. Our results show a proof-of-concept that a QA-based RBM can be trained on a 64-bit binary dataset. The illustrative example suggests the possibility to migrate many practical classification problems to QA-based techniques. Further, we show that synthetic data generated from a RBM can be used to balance the original dataset.
\end{abstract}


%
\IEEEpeerreviewmaketitle

\section{Introduction}
\IEEEPARstart{N}{etworks} have revolutionized our lives through various purposes including email, file transfer, web search, e-commerce, online banking, monetary transaction, education, collaboration, social networking, etc. The more we depend on it and use it, the more we expose ourselves to serious security risks. The internet is an insecure medium of communication. Any device connected to the internet is vulnerable. Cybersecurity is safety against cyber-attacks. Cyber-attacks are launched by hackers to gain unauthorized access or steal important data. The estimated total damage caused by global cybercrime has increased from \$300 billion in 2013  to \$945 billion in 2020 \cite{Smith-2020}.  Financial loss from cybercrime is likely to increase in the coming years. Therefore, it is crucial to monitor dataflow in any network, and there is a need for robust software and devices that protect users from online security threats.

In this work, we investigate a restricted Boltzmann machine (RBM) coupled with quantum machine learning for a cybersecurity application.  The application of quantum computing in machine learning is a promising technique, even with quantum computers currently being in an early stage of technological development.  This paper is a first approach of implementing for network intrusion detection an analysis engine on a quantum computing device. A RBM is a generative model, which can be used  to model the underlying probability distribution of a dataset. In addition to classifying data points, RBMs can also generate a new synthetic dataset.  Despite the importance of the RBMs, only a few researchers have used RBMs for cybersecurity applications. Fiore \emph{et~al.} \cite{fiore2013network} used discriminative RBM for network anomaly detection applications.  They showed that the performance of a model suffers when it is tested in a network different from the network that was used to obtain the training data.  Aldwairi \emph{et~al.} \cite{aldwairi2018evaluation} trained a RBM on the  ISCX 2012 dataset using contrastive divergence (CD) and persistent contrastive divergence (PCD). Their model showed a percentage classification accuracy of 88.6\% using CD and 89\% for PCD. Alom \emph{et~al.} \cite{alom2015intrusion} applied a deep belief network (DBN) on the  NSL-KDD D'99 intrusion detection dataset.  They were able to get a classification accuracy of 97.5\% just by using 40\%  of  the dataset. A DBN model is composed of multiple layers of trained RBMs, weights are fine-tuned by performing backpropagation in the final step of the model training.  Salama \emph{et~al.} \cite{salama2011hybrid} used a  DBN+SVM hybrid scheme for intrusion detection. They used a DBN for dimensionality reduction (from 41 to 5 features) and SVM for classification. The model was trained on the  NSL-KDD dataset.   Li \emph{et~al.} \cite{li2015hybrid} trained a hybrid model on 10\% KDDCUP'99 dataset. An autoencoder was used to reduce the dimensionality of the dataset and a DBN for classification.   Alrawashdeh \emph{et~al.} \cite{alrawashdeh2016toward} trained a DBN on the  KDDCUP'99 dataset.  Their model outperformed the model  by  Salama \emph{et~al.}  \cite{salama2011hybrid}   and  Li \emph{et~al.} \cite{li2015hybrid} both in speed and accuracy. 

We have used a quantum annealer from D-Wave to train RBMs for intrusion detection applications and compared the performance to RBMs trained with contrastive divergence. Quantum annealers are based on adiabatic quantum annealing (QA), which is a powerful technique for optimization and sampling applications \cite{mizel2007simple, farhi2000quantum, aharonov:2007, Kadowaki:1998, FINNILA1994343}. There are two main problems associated with the use of machine learning techniques for intrusion detection. The first problem is related to transferability and generalizability of the model, a model trained on a dataset performs poorly when tested on other datasets. The second problem is associated with the imbalanced nature of the cybersecurity dataset where the attack instances are outnumbered by benign instances, which makes detection of an attack like looking for a needle in a haystack. Quantum computing holds the promise to address  these problems. A QA-trained RBM can effectively learn patterns without overfitting a dataset. Further, synthetic data from a RBM can be used to balance the original dataset. Our work is a step towards that goal.  The D-Wave 2000Q adiabatic quantum computer has been used by several researchers for machine learning applications such as classification, regression, and clustering. Date \emph{et~al.} \cite{potok2020adiabatic} used a quantum annealer for implementing linear regression.  Willsch \emph{et~al.} \cite{Willsch2020} introduced a method to train support vector machines (SVMs) on a D-Wave 2000Q quantum annealer and compared its performance with classically trained SVMs. Kumar \emph{et~al.} \cite{kumar} used quantum annealing to carry out the minimization of the clustering objective function. They implemented two clustering algorithms and compared their results with well-known k-mean clustering. Das \emph{et~al.} \cite{das2019track} used a D-Wave to implement a clustering algorithm for the clustering of charged particle tracks for a hadron collider experiment. Arthur \emph{et~al.} \cite{arthur2020} used the D-Wave 2000Q adiabatic quantum computer to train the balanced k-means clustering model. They compared the results with classical k-means and classical balanced k-means. Kais \emph{et~al.} have used D-Wave's quantum annealer for prime factorization and electronic structure calculation of molecular systems \cite{Kais2018a, Kais:2018}. Adachi \emph{et~al.} \cite{adachi2015application} trained RBMs using a quantum annealer for a deep belief network (DBN) on a scaled-down MNIST dataset consisting of 32-bit length binary records. They showed that their model required fewer iterations than CD-based DBN training. Benedetti \emph{et~al.} \cite{PhysRevA.94.022308} used quantum annealing to train a RBM on a 16-bit binary bars \& stripes dataset.  Koshka \emph{et~al.} \cite{koshka2020toward, k91} trained a RBM using contrastive divergence and compared the samples obtained from Markov chain Monte Carlo (MCMC) and QA. For QA, the CD trained RBM was embedded onto the D-Wave, and sampling was performed. It was found that the QA based sampling revealed regions of the configuration space that were regularly missed by the MCMC based sampling, especially at medium to high energy (i.e., states of medium to low probability). 
Recently, Dixit \emph{et~al.} \cite{dixit2020training} trained a RBM using the D-Wave 2000Q quantum annealer for classification and image reconstruction applications. They used a 64-bit bars \& stripes dataset in their work. 

The D-Wave 2000Q has around 2000 qubits.  D-Wave’s recently introduced  machine `Advantage’ comprises 5000 qubits. The number of qubits of a quantum annealer is a major factor that  determines the size of a dataset that can be investigated. Sometimes the number of features of a large dataset can be reduced by finding a dense representation. Caldeira \emph{et~al.} \cite{caldeira2020} used PCA to reduce the number of features in the dataset. They used a QA-trained RBM for galaxy morphology classification. Sleeman \emph{et~al.} \cite{sleeman2020hybrid} used an autoencoder to obtain a dense reperesentation of their dataset. They were able to show  nearly a 22-fold compression factor of grayscale 28 x 28 sized images to binary 6 x 6 sized images.  They trained a QA-based RBM on the MNIST and the MNIST Fashion datasets.

\section{Contribution}

Cybersecurity is one of the key areas where the failure of detection systems can result in privacy intrusion, financial losses, and system shutdowns. Our goal is to train the RBM using a quantum annealer, to help explore quantum effects for faster training and to learn patterns efficiently.  Given that network data is usually imbalanced, we seek to obtain synthetic samples generated by a RBM to provide rich information into the distribution from which attack samples are generated.  This should enable classifiers to better train on and detect intrusions. There are two main objectives of this work. First, train a RBM using quantum annealing on a cybersecurity dataset (ISCX). Second, use a RBM to generate synthetic data to balance the cybersecurity data.

First, we show RBMs can be trained using a quantum annealer on a cybersecurity dataset. Conventional methods for RBM training such as CD and PCD are slow. They require many Gibbs cycles to train a RBM. Further, the CD does not estimate the correct gradient of log-likelihood \cite{hinton:2002}. RBM training using a quantum annealer is fundamentally different than existing methods. A quantum annealer exploits quantum effects like superposition and tunneling to find better low energy solutions. This could be particularly useful for intrusion detection applications where classifiers often show poor precision and accuracy. We believe that this is the first work that uses a QA-trained RBM for intrusion detection applications.

The second objective is to show that synthetic data from  a RBM can be used to balance a cybersecurity dataset. Cybersecurity datasets often have a lower number of attack records. However, most of the machine learning techniques require a balanced dataset. A bias towards the majority class results if the dataset is not balanced. A commonly used method to balance a dataset is SMOTE (Synthetic Minority Over-sampling Technique) \cite{chawla2002smote}. The SMOTE algorithm basically works by finding the k-nearest neighbor of a data point in the feature space of the minority class. Then a synthetic data point is obtained by interpolation between the data point and one of the k-neighbors. Generally, this  interpolation is performed based on a random number between 0 and 1. This process is repeated until the required number of synthetic data records is obtained. Several modifications and extensions of SMOTE have been made since its proposal \cite{fernandez2018smote}. Several investigators have used SMOTE for cybersecurity applications \cite{ma2020aesmote, yan2017novel, su2018research, tesfahun2013intrusion, ahsan2018smote}. A trained RBM can be used to generate synthetic data records. An advantage of  using a RBM is that the synthetic data from it follows the probability distribution of the training dataset. However, synthetic data from SMOTE might not follow the distribution of the training data. In this work, we use QA-trained as well as CD-trained RBMs to generate synthetic data. This synthetic dataset  is then used to balance the original dataset. 

Herein, we propose two schemes to balance the cybersecurity dataset. The first scheme is based on the under-sampling of benign records. In the second scheme, oversampling of the attack class is used. Synthetic data has been generated from a RBM to balance the training dataset. RBMs are trained on a benchmark intrusion detection dataset known as ISCX \cite{shiravi2012toward}.

\section{Methods} 
\subsection{Restricted Boltzmann Machine (RBM)}

A RBM is an undirected probabilistic graphical model consisting of a layer of visible variables and a single layer of  latent or hidden variables. Each variable is connected to every variable  in the opposite layer, but  connections between the variables in the same layer are not allowed. Let the visible and hidden layers be composed of $N$ and  $M$variables denoted as  $\{ v_1, v_2, ....,v_N\}$ and $\{ h_1, h_2, ....,h_M\}$, respectively. We collectively refer to the visible layer with the vector $v$ and the  hidden layer as $h$. The RBM is an energy-based model with the joint probability distribution specified by its energy function:

\begin{equation}\label{eq:3}
P(v,h)=\frac{1}{Z} e^{-E(v,h)}, \space \space  Z=\sum_v \sum_h e^{-E(v,h)}.
\end{equation}
$Z$ is the normalization constant known as the partition function.
The energy function is defined as:
\begin{equation}\label{eq:2}
E(v,h)=-b^Tv-c^Th-h^TWv,
\end{equation}
where $b$ and $c$ are bias vectors at the visible and hidden layer, respectively; $W$ is a weight matrix composed of $w_{ij}$ elements.

\subsection{Conditional Distribution}

The probability of getting a vector $h$ at the hidden layer given a vector $v$ at the visible layer is:

\begin{equation}\label{eq:8}
P(h|v)= \frac{P(v,h)}{P(v)}
\end{equation}

where $P(v)$ is given by the following expression:

\begin{equation}\label{eq:9}
P(v)= \frac{\sum_h e^{-E(v,h)}}{Z}.
\end{equation}

Using expression $P(v,h)$ from Eq. \ref{eq:3}, we get:

\begin{equation}\label{eq:10}
P(h|v)= \frac{\exp\{{\sum_j c_j h_j + \sum_j (v^T W)_j h_j}\} }{Z^\prime},
\end{equation}

where
\begin{equation}\label{eq:11}
 Z^\prime = \sum_h \exp(c^Th + h^TWv).
\end{equation}

\begin{equation}\label{eq:12}
P(h|v)= \frac{1}{Z^\prime} \displaystyle\prod_{j} \exp\{{c_j h_j + (v^T W)_j h_j}\}.  
\end{equation}

Let's denote

\begin{equation}\label{eq:13}
\widetilde{P}(h_j|v) = \exp{\{c_j h_j + (v^T W)_j h_j\}}
\end{equation}

The probability to find an individual variable in the hidden layer, $h_j = 1 $  is:
\begin{equation}\label{eq:14}
\begin{split}
P(h_j = 1|v) & =\frac{\widetilde{P}(h_j=1|v)}{\widetilde{P}(h_j=0|v) + \widetilde{P}(h_j=1|v)} \\
& = \frac{\exp\{c_j + (v^T W)_j \}}{1+\exp\{c_j + (v^T W)_j \}}
\end{split}
\end{equation}

Thus, the individual hidden activation probability is given by:

\begin{equation}\label{eq:15}
P(h_j = 1|v) = \sigma\Big(c_j + (v^T W)_j \Big),
\end{equation}

\noindent where $\sigma$ is the logistic function. Similarly, the activation probability of a visible variable conditioned on a hidden vector $h$  is given by:

\begin{equation}\label{eq:16}
P(v_i = 1|h) = \sigma\Big(b_i + (h^T W)_i \Big).
\end{equation}

\subsection{RBM Training}

A RBM is trained by maximizing the likelihood of the training data. The log-likelihood is given by:

\begin{equation}\label{eq:17}
\begin{split}
l(W,b,c) & = \sum_{t=1}^{N} \log P\Big(v^{(t)}\Big) \\
& = \sum_{t=1}^{N} \log \sum_{h} P\Big(v^{(t)},h\Big),
\end{split}
\end{equation}
where $N$ is the number of records in the training dataset and $v^{(t)}$ is a sample from the training dataset. 
\begin{equation}\label{eq:18}
\begin{split}
l(W,b,c) & = \sum_{t=1}^{N} \log \sum_{h} e^{-E(v^{(t)},h)} \\
&- N \cdot \log \sum_{v,h} e^{-E(v,h)}.
\end{split}
\end{equation}

Denote $ \theta=\{W, b, c\} $. The gradient of the log-likelihood is given by:

\begin{equation}\label{eq:19}
\begin{split}
\nabla_{\theta} l(\theta)  & = \sum_{t=1}^{N} \frac{\sum_h e^{-E(v^{(t)},h)} \nabla_\theta (-E(v^{(t)},h))}{\sum_h e^{-E(v^{(t)},h)}} \\
&  - N \cdot \frac{\sum_{v,h} e^{-E(v,h)} \nabla_\theta (-E(v,h))}{\sum_{v,h} e^{-E(v,h)}} 
\end{split}
\end{equation}

\begin{equation}\label{eq:20}
\begin{split}
\nabla_{\theta} l(\theta) & = \sum_{t=1}^{N} \langle\nabla_\theta(-E(v^{(t)},h))\rangle_{P(h \mid v^{(t)})} \\
& - N \cdot \langle\nabla_\theta(-E(v,h))\rangle_{P(v, h)},
\end{split}
\end{equation}

\noindent where $\langle \cdot \rangle_{P(v, h)}$ is the expectation value with respect to the distribution $P(v,h)$.  The gradient with respect to $\theta$ can also be expressed in terms of its components:

\begin{equation}\label{eq:21}
\nabla_{w} l = \frac{1}{N}\sum_{t=1}^{N} \langle v^{(t)}\cdot h\rangle_{P(h \mid v^{(t)})} -  \langle v\cdot h\rangle_{P(v, h)}
\end{equation}

\begin{equation}\label{eq:22}
\nabla_{b} l = \frac{1}{N}\sum_{t=1}^{N} \langle v^{(t)}\rangle_{P(h \mid v^{(t)})} -  \langle v\rangle_{P(v, h)}
\end{equation}

\begin{equation}\label{eq:23}
\nabla_{c} l = \frac{1}{N}\sum_{t=1}^{N} \langle h\rangle_{P(h \mid v^{(t)})} -  \langle h\rangle_{P(v, h)}.
\end{equation}

The first term in Eq. \ref{eq:20}  is a data-dependent term. It can be exactly calculated using a training vector $ v^{(t)}$ and a hidden vector $h$. Given  $ v^{(t)}$,  the vector $h$ can be calculated using Eq. \ref{eq:15}. The second term is a model-dependent term. Getting samples for the second term is difficult. The Contrastive Divergence (CD) is the most commonly used algorithm to determine the model-dependent term. In CD, a training vector is applied to the visible layer. Then the binary states of the hidden units are computed in parallel using  Eq. \ref{eq:15}. The states of the units on the visible layer are reconstructed using $h$  via Eq. \ref{eq:16}. Finally, the reconstructed $v$ is used to find a new $h$ on the hidden layer. During the RBM training, the change in model parameters is given as: 
\begin{equation}\label{eq:24}
  \theta_j^{new} = \theta_j^{old} + \epsilon \cdot \nabla_{\theta_j} l(\theta_j)
\end{equation}
where $\epsilon$ is the learning rate.  

The learning works well even though CD only crudely approximates  the gradient of the log probability of the training data. Sutskever  \emph{et~al.}  \cite{sutskever2010} have shown that the contrastive divergence does not estimate the gradient of the log-likelihood. An effective method for RBM training is still not known. It has been shown by several researchers that a RBM can be trained using  samples drawn from the D-Wave quantum annealer \cite{adachi2015application} \cite{PhysRevA.94.022308} \cite{caldeira2020} \cite{dixit2020training}. The first term of the gradient of the log-likelihood is estimated using the procedure explained earlier. The second term which is the model-dependent term is calculated in the following way.  First, a RBM is embedded on to a quantum annealer, then quantum annealing is performed. The samples obtained from quantum annealing are used to compute the second term. The samples from a quantum annealer operating at a temperature $T$ is qualitatively similar to a probability distribution given by $\exp(\frac{-E(v,h)}{kT})$. However, to compute the model-dependent term we need samples from a distribution $\exp({-E(v,h)})$ (Eq. \ref{eq:20}).   To address this problem we scale the energy by a hyperparameter $S$, such that for the model-dependent term, we sample from the $\exp({\frac{-E(v,h)}{SkT}})$ distribution.  Here, $S$ is a  hyperparameter, which is determined by a manual search. The optimal condition corresponds to the case when $SkT=1$. We keep $S$ fixed during the entire RBM training process. However, the temperature $T$ generally changes. This mismatch between $S$ and $T$ might result in suboptimal RBM training. An efficient way to compute $T$ at each training step is still not discovered.  It should be noted as the RBM training starts with random weights and biases,  samples from the D-Wave are not expected to show a Boltzmann distribution, however, as the training progresses the underlying probability distribution moves toward the Boltzmann distribution. RBM training using CD-1 and QA is summarized in Algorithm \ref{algo:1} and Algorithm \ref{algo:2}, respectively. The two methods differ only in the manner the model-dependent term is estimated.

\subsection{The D-Wave Quantum Annealer}
 
To formulate a problem for the D-Wave, one needs to transform the problem into Ising form given by:

\begin{equation}\label{eq:25}
E(s|h,J) =  \sum_{i=1}^{N} h_i s_i  + \sum_{i<j}^{N} J_{ij}s_i s_j;    \:       s_i \in \{-1,+1\}.
\end{equation}

This is an objective function of $N$ variables $s = [s_1, s_2, ..., s_N ]$ corresponding to physical Ising spins, where $h_i$ are the biases and $J_{ij}$ the couplings between spins. 

The energy of a RBM model given by Eq. \ref{eq:2}, has a form similar to Eq. \ref{eq:25}. The weights and biases of a RBM which is trained using a binary dataset, $\{0, 1\}$ states can be converted to use $\{-1,1\}$ states via the mapping \cite{Dumoulin}:

\begin{equation}\label{eq:26}
b_i^{\prime} = \frac{b_i}{2} + \frac{\sum_{j} W_{ij}}{4}
\end{equation}
 
\begin{equation}\label{eq:27}
c_i^{\prime} = \frac{c_i}{2} + \frac{\sum_{j} W_{ij}}{4}
\end{equation}

\begin{equation}\label{eq:28}
W^{\prime} = \frac{W}{4}.
\end{equation}

These weights and biases can be used to embed a RBM onto the D-Wave machine. After executing  quantum annealing, solutions can be sampled. We set the anneal time to $20\mu s$  for each anneal. The resulting bipolar samples may be converted to a binary sample simply by replacing all instances of -1 with 0. 

\begin{figure*}[h!]
\centering
\includegraphics[scale=0.65]{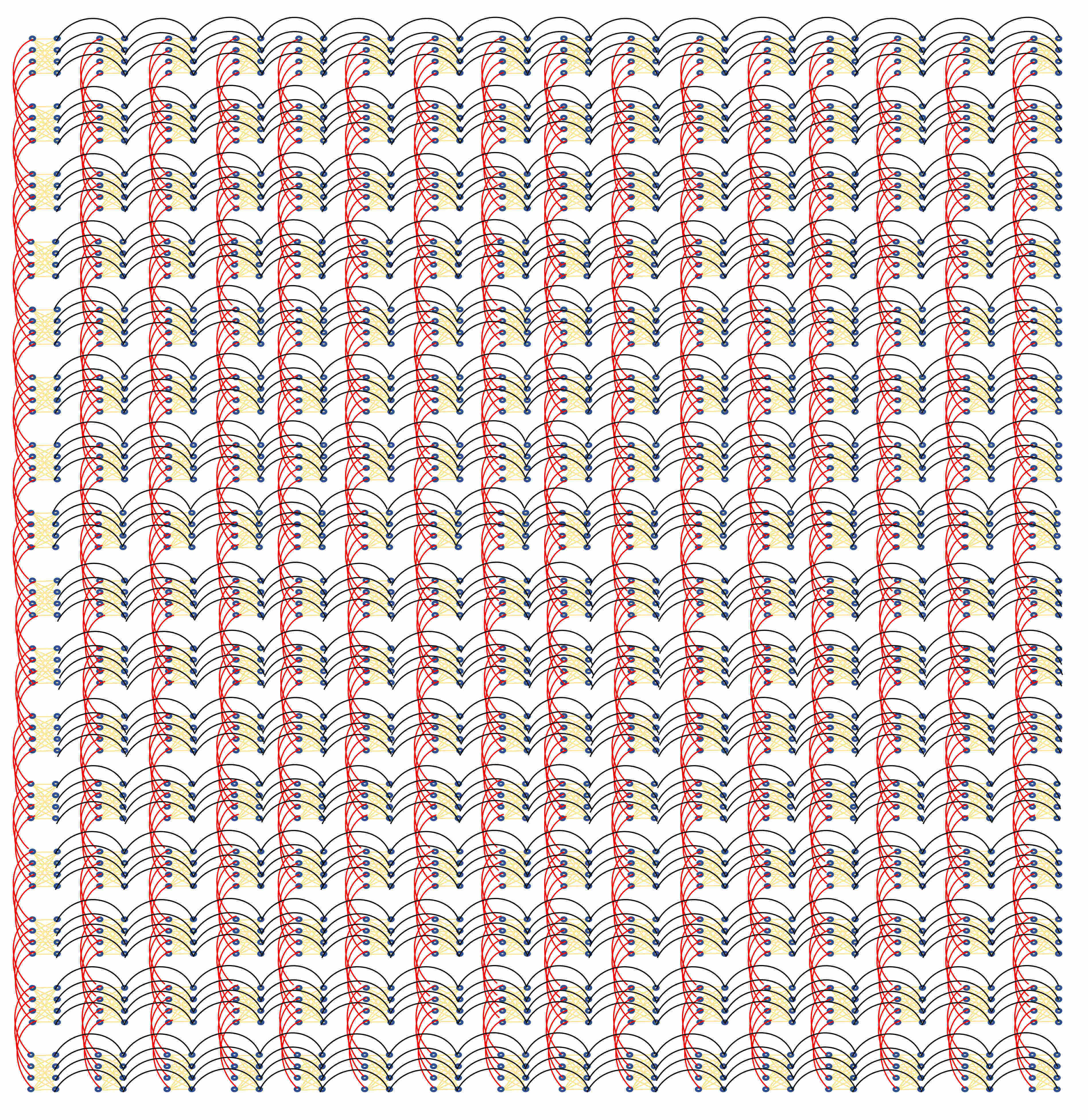}
\caption{\label{fig:chain} Minor-Embedding a RBM with 64 visible and 64 hidden units on the D-Wave 2000Q. Each visible (hidden) unit is made by forming a vertical (horizontal) chain of 16 physical qubits shown in red (black).} 
\end{figure*}

The D-Wave 2000Q quantum annealer has 2048 qubits arranged in $16 \times16$ unit cells forming a C16 Chimera graph. Each unit cell is composed of 8 qubits connected in a bipartite graph. Each qubit is connected to four other qubits of the same unit cell and two qubits of two different unit cells. One can embed a fully connected RBM of 64 visible and 64 hidden units on a C16 Chimera graph as shown in Fig. \ref{fig:chain}.  In this embedding, each RBM unit is represented by a chain of 16 physical qubits. If we look at the arrangement of qubits, we note that 16 qubits can be combined by forming a vertical chain. Each vertical chain forms one visible unit. Similarly, 16 horizontal qubits can be linked together to form a hidden unit. In Fig. \ref{fig:chain}, the vertical chains are shown in red, while the horizontal chains are in black. There are 64 vertical and 64 horizontal chains which represent 64 visible units and 64 hidden units of the RBM. If some of the qubits in the D-Wave QPU are missing or not working, then the length of the qubit chain forming a RBM unit will be shorter. In that case, the RBM will be not fully connected, that is some connections between visible and hidden units will be missing. Fortunately, in the D-Wave QPU, only a few qubits are missing which does not  seem to affect the performance of the RBMs.  This embedding has also been used in our previous work \cite{dixit2020training}. A similar bipartite embedding has  been demonstrated by other researchers \cite{Humble:2018, koshka2017determination, koshka2018comparison}.
For the lattice with almost no missing qubits and couplings, this embedding is close to optimal for a Chimera graph. We have maximally made use of the D-Wave 2000Q to allow for 64 hidden qubits and visible qubits that are two way fully connected to each other. Using any other embedding would result in a small size of the feature space and hence is not preferred. The newer machine (Advantage) has over 5000 qubits and additional graph connections, which should allow for an extended feature space size where one could do a comparison of the performance of different embedding schemes for this dataset.

\begin{algorithm}
\caption{Optimization of learning parameters using CD-1}\label{algo:1}
\begin{algorithmic}[1]

\State $\mathit{\epsilon} \gets \mathit{\textit{learning rate}}$     \Comment{$\epsilon$, is the step size, a small positive number.}
\State $b$, $c$, $W$ $\gets$ $\textit{random number}$ \Comment{Initialize with small normally distributed random numbers.}
\While {\textit{not converged}} 
  \State Sample a minibatch of $m$ examples $ \{x^{(1)},...,x^{(m)}\} $ from the training set             
  \State $V \gets  \{x^{(1)},...,x^{(m)}\} $ 
  \State $H \gets \sigma(c + VW)$ \Comment{$\sigma$ is the logistic function}                                 
  \State $V^\prime \gets \sigma(b + HW^T)$
  \State $H^\prime \gets \sigma(c + V^\prime W)$
  \State $W \gets W + \epsilon \frac{\big(  VH - V^\prime H^\prime \big)}{m}$ \Comment{updates $W$}
  \State $b \gets b + \epsilon \frac{\big(  sum(V) - sum(V^\prime) \big)}{m}$ \Comment{updates $b$}
  \State $c \gets c + \epsilon \frac{\big(  sum(H) - sum(H^\prime) \big)}{m}$ \Comment{updates $c$}
\EndWhile
\State \textbf{end}
\end{algorithmic}
\end{algorithm}

\begin{algorithm}
\caption{Optimization of learning parameters using quantum annealing}\label{algo:2}
\begin{algorithmic}[1]

\State $\mathit{\epsilon} \gets \mathit{\textit{learning rate}}$     \Comment{$\epsilon$, is the step size, a small positive number.}
\State $b$, $c$, $W$ $\gets$ $\textit{random number}$ \Comment{Initialize with small normally distributed random numbers.}
\While {\textit{not converged}} 
  \State Sample a minibatch of m examples $ \{x^{(1)},...,x^{(m)}\} $ from the training set             
  \State $V \gets  \{x^{(1)},...,x^{(m)}\} $ 
  \State $H \gets \sigma(c + VW)$  \Comment{$\sigma$ is the  logistic function}
  \State $\{h,J\} \gets \{ b, c, W\}$
  \State $(V^\prime, H^\prime  ) \gets \textit{quantum annealing}(h, J, S)$
  \State $W \gets W + \epsilon \frac{\big(  VH - V^\prime H^\prime \big)}{m}$ \Comment{updates $W$}
  \State $b \gets b + \epsilon \frac{\big(  sum(V) - sum(V^\prime) \big)}{m}$ \Comment{updates $b$}
  \State $c \gets c + \epsilon \frac{\big(  sum(H) - sum(H^\prime) \big)}{m}$ \Comment{updates $c$}
\EndWhile
\State \textbf{end}
\end{algorithmic}
\end{algorithm}

\subsection{Evaluation Metrics}

To compare and quantify the performance of different methods,  metrics  based on a confusion matrix are used. For a binary classification problem that has two classes namely `positive' and `negative' important metrics for model evaluation are:

\begin{equation}\label{eq:29}
A_{tot} = \frac{TP+TN }{TP+TN+FP+FN} \times 100,
\end{equation}

\begin{equation}\label{eq:30}
A_P = \frac{TP}{TP+FP} \times 100,
\end{equation}

\begin{equation}\label{eq:31}
\textit{Precision} = \frac{\textit{TP}}{\textit{TP+FP}}
\end{equation}
\begin{equation}\label{eq:32}
\textit{Recall} = \frac{\textit{TP}}{\textit{TP+FN}}
\end{equation}
\begin{equation}\label{eq:33}
F_1 \textit{score} = 2 \times \frac{\textit{Precision} \times \textit{Recall}}{\textit{Precision + Recall}}
\end{equation}

\noindent where TP (true positive) and FP (false positive) are the number of correctly and incorrectly predicted observations of class `positive', respectively. Similarly, TN (true negative) and FN (false negative) are the  number of correctly and incorrectly predicted observations of class \lq negative', respectively. $A_{tot}$ is the total percentage of classification accuracy; $A_P$ is the percentage of classification accuracy of the class \lq positive'.  Precision is the ability of the model not to predict the label of a sample of a class incorrectly, while recall is the ability of the model to correctly predict all the samples of a class correctly. The  $F_1$ score is the harmonic mean of precision and recall. A robust classifier will have a high value of the $F_1$ score. Precision, recall, $F_1$ score, and percentage accuracy are used as metrics to evaluate models.

\subsection{Material Setup}
 The D-Wave 2000Q quantum annealer has been used to obtain samples of training QA-based RBM. The D-Wave operates at a temperature that is fixed based on the training results. The temperature corresponds to an effective scaling of parameters that are supplied as coupling weights and biases to the machine. 

 For training CD-based RBM, a personal computer has been used. In-house codes were developed to implement RBM training using CD and to obtain samples from the quantum annealer. MATLAB programming language is used.  MATLAB codes are also developed for classification and to generate synthetic data.  To implement popular classification methods namely: Neural Networks, K-nearest neighbor, Support Vector Machine, Decision tree, and Naive Bayes, machine learning library for python programming language `scikit-learn' \cite{scikit-learn} has been  used. All the classifiers are trained on 62-bit binary input and 2-bit output data. A neural network with five layers has been used. Layer 1 has 62 nodes. There are three hidden layers, each with 8 nodes. The output layer has 2 nodes. The `relu' activation function and \lq adam' solver have been used for training. To implement the K-nearest neighbor classifier  neighbors is set 3, other parameters are set to their default values. The decision tree classifier has a max depth set to 5, other values are set to the  default. To implement support vector machine and Naive Bayes classifiers, SVC and GaussianNB classifiers of `scikit-learn' have been used. For these two classifiers, all the parameters are set to their default values.

\subsection{Dataset}

This study investigates a cybersecurity  benchmark dataset known as `ISCX IDS dataset 2012’. We will call it ISCX, for brevity. The ISCX  is one of the publicly available datasets on the website of the Canadian Institute of Cybersecurity at https://www.unb.ca/cic/datasets/index.html. The ISCX consists of seven days of network activity. There are two main classes namely `benign’ and `attack’. For more details and the underlying approach that is used to generate this dataset, see \cite{shiravi2012toward}.  The preprocessing of the data consists of the conversion of variables from categorical  to numerical, dimensionality reduction, and binarization of numerical data. All these steps are necessary to build a classifier based on a  QA trained RBM.  Finally, the dataset is binarized by using a supervised discretization filter implemented in Weka \cite{witten2002data}.  As we discussed earlier, one can embed a RBM onto the D-Wave 2000Q with 64 visible and 64 hidden units. Therefore, we set the total number of columns in the binarized dataset to be 64. There are 62 binary features in the dataset and the last two columns are the target variables. When the last two bits are 01, it indicates a benign instance; while 10 indicates an attack. If the last two bits are either a 00 or 11, it indicate an indeterminate case. Thus, the possibility that a random guess could be correct is 25\%, and keeping two bits for the target variable helps to prepare a more robust machine learning model as compared to the case where one bit is used as a target variable. 

\section{Results}

\begin{figure*}[h!]
\begin{center}
\includegraphics[scale=0.9]{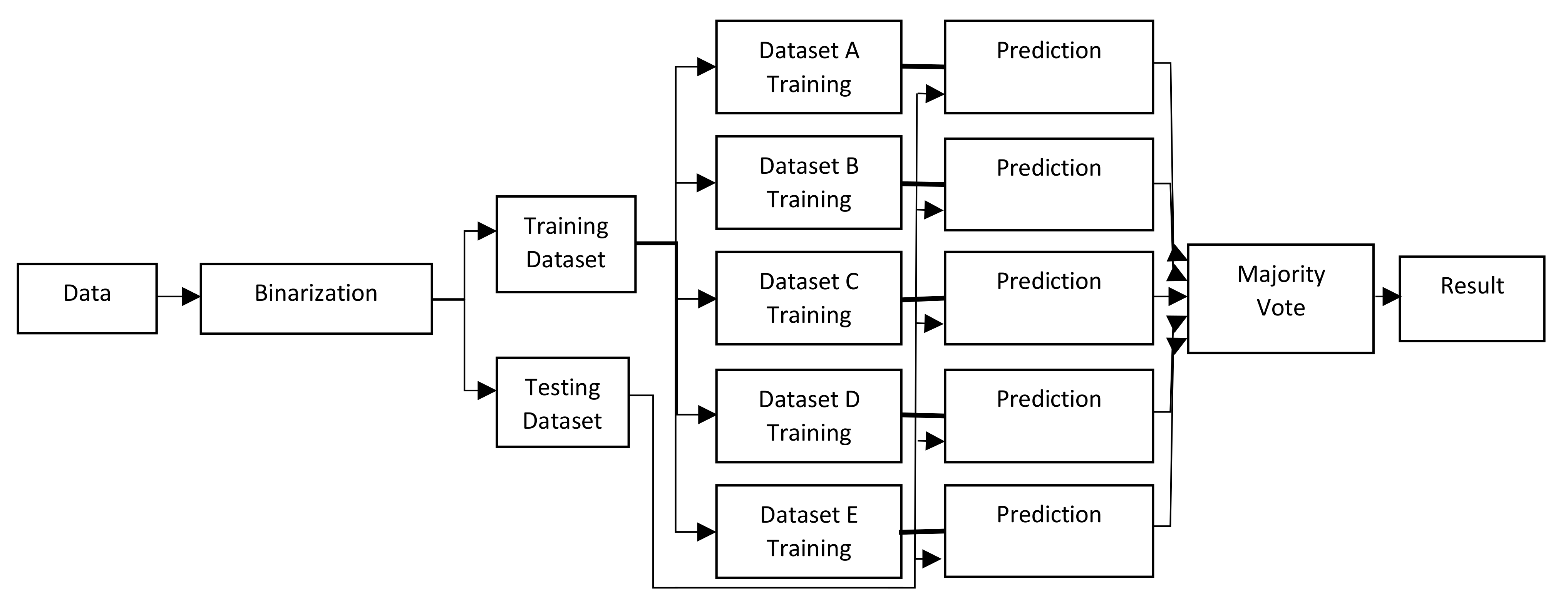}
\end{center}
\caption{\label{fig:S1} Scheme 1: Flowchart for intrusion detection using an imbalanced dataset. The training dataset is divided into five balanced datasets which are subsequently used to train five classifiers. The final classification result is obtained by a majority voting.} 
\end{figure*}

The dataset that was obtained after binarization of the original dataset had 137584 instances. However, it was found that most of the records were repeated. The dataset was further modified, and only unique records were retained. There were 25230 unique benign and 4917 unique attack records. Training and test datasets are formed with these records. 
The test dataset comprises 500 attack and 500 benign records.  The remaining 29147 unique records are used in the training dataset. We trained a RBM on the training dataset. The classification accuracies for the attack and benign classes are found to be 42\% and 97\%, respectively.  These accuracies are estimated on the test dataset. The lower accuracy for the attack class could be attributed to the fact that there are a significantly higher number of benign instances in the dataset compared to the number of attack instances. The attack records constitute only 14.1\% of the total dataset, however, ideally, there should be 50\% records of each class. The problem of an imbalanced dataset is commonly seen in cybersecurity datasets; attack records form a rarer class. Machine learning algorithms show the best results when the number of observations in each class is almost similar. Thus, an imbalanced dataset leads to a poor classification performance of the model. This imbalanced dataset is also investigated using other classification methods, and the results are presented in Table \ref{tab:models}. To tackle the problem of an imbalanced dataset we propose two schemes. In the first scheme, we use undersampling of the benign class, while in the second scheme a RBM has been used  to generate instances in order to balance the training dataset. These schemes are discussed in  detail in the following sections.

\subsection{Scheme 1: Balancing training data by undersampling of benign records}

\begin{table*}[h!]
  \caption{\label{tab:cd_vs_qa} The original training dataset is divided into five small datasets (A, B, C, D, and E). These datasets are used to train five RBMs using contrastive divergence (CD-1) and quantum annealing (QA). Classification accuracies for benign and attack classes, as well as total accuracy, are presented for each dataset. Values are evaluated on the testing data. A majority vote is  performed on the results obtained from five RBMs. }
  \centering
    \begin{tabular}{|c|c|cc|cc|cc|}
     \hline
\noalign{\smallskip}
     \multirow{2}{*}{dataset}   &  \multirow{2}{*}{No. of records}  &\multicolumn{2}{c|}{Accuracy, $A_{benign}$} & 
      \multicolumn{2}{c|}{Accuracy, $A_{attack}$ }   &   \multicolumn{2}{c|}{Total Accuracy, $A_{tot}$ }\\
\cline{3-4}\cline{5-6} \cline{7-8}
\noalign{\smallskip}
      &  & CD-1  (\%) & QA  (\%)  &  CD-1  (\%)  & QA  (\%) & CD-1  (\%)  & QA (\%)\\
      \hline
\noalign{\smallskip}
 A & 6900 & 92.17 & 82.60 & 88.75  & 68.30 & 91.59 & 80.19\\
 B & 6900 & 88.30 & 71.76 & 88.82  & 72.60 & 88.39 & 71.90\\
 C & 6900 & 90.11 & 69.04 & 89.23  & 83.01 & 89.96 & 76.03\\
 D & 6900 & 90.03 & 67.65 & 90.52  & 72.73 & 90.11 & 68.51\\
 E & 8400 & 91.94 & 77.05 & 87.39  &  59.24 & 91.17 & 74.05\\
\hline
\noalign{\smallskip}
Average & -- & 90.51 & 73.62 & 88.94  & 71.18 & 90.24 & 74.14\\
Standard deviation & -- &  1.59 & 6.17 & 1.12  & 8.59 & 1.25 & 4.38\\
Majority Vote & -- & 96.17 & 74.46 & 93.25  & 85.62 & 95.68 & 80.04\\
\hline
      \end{tabular}
\end{table*}

Scheme 1 is illustrated in Fig. \ref{fig:S1}. In this approach, the binarized dataset is divided into training and testing datasets. The training dataset is composed of 21450 records (Benign=18000, Attack=3450), while the test data contains 8697 records (Benign=7230, Attack=1467). Thus, the original binarized dataset is divided into training and testing datasets in a  ratio of $\approx70\%:30\%$. The training dataset is further divided into five smaller datasets namely A, B, C, D, and E. The total number of benign records in the training dataset is divided into five datasets as $18000=3450+3450+3450+3450+4200$.  Thus, each sub-dataset has unique benign records. There are 3450 attack records in the training dataset, we add the same 3450 attack records to each sub-datasets. Sub-dataset E has 4200 (=3450+750) attack records, 750 of which are  repeated. Thus, each sub-dataset contains  an equal number of instances of both classes. Five RBM models are trained on these five datasets. These trained RBMs models are used to make predictions on the testing dataset. Predictions from the five RBM models are collected and a majority vote rule  has been performed to obtain a final result. Two different methods, contrastive divergence (CD-1) and quantum annealing (QA), are employed to train the RBMs. In Table \ref{tab:cd_vs_qa} we show the average classification accuracy of the benign class is 90.51\% and that of the attack class is  88.94\%. The total accuracy is 90.24\%.  On using the majority vote on the results obtained from five different RBMs, the classification accuracies with which the benign and attack classes can be predicted, and total accuracy have been found to be 96.17\%, 93.25\%, and 95.68\%, respectively. In the case where RBMs are trained with quantum annealing, the average classification accuracy of the benign and attack classes, and total accuracy are 73.62\% and 71.18\%, and 74.14\%, respectively. On applying the majority vote on the results from five trained RBMs, the average classification accuracy for the benign, attack classes, and total accuracy are found to be 74.46\%, 85.62\%, and 80.04\%, respectively. Thus, in the case of CD-1 as well as QA, we note an improvement in accuracy when the majority vote is applied.  Table \ref{tab:cd_vs_qa} also compares the performances of RBMs trained using CD-1 and QA methods. Using the majority vote the total accuracy with CD-1 and QA methods are found to be 95.68\% and 80.04\%, respectively. If we consider the results from the individual models, for example, the RBM model trained on sub-dataset A. Dataset A is comprises of just 3450 attack and 3450 benign records, but the classification accuracy of the RBM is better than the case when the training dataset was imbalanced (Table \ref{tab:models}).
The contrastive divergence being a state-of-the-art method for RBM training, a better performance of a CD trained RBM is expected. While CD-1 is a popular and effective method for RBM training, QA for RBM training has not been substantially explored. Considering the prevailing noise and error-prone nature of the existing quantum machine a classification accuracy of 80.04\% seems to be satisfactory. Our goal here is to show a proof-of-concept that RBM can be trained using quantum annealing on a 64-bit binary dataset. RBM training using QA can be improved by optimizing D-Wave annealing parameters like anneal time, chain length, etc. Further, an efficient way to calculate the quantum annealer's effective temperature can also improve QA-based RBM training.

\subsection{Scheme 2: Balancing training dataset with synthetic data}

A dataset is said to be imbalanced if the number of observations in each class is not proportionate. Generally, when we deal with a cybersecurity dataset, we face the problem of a lower number of attack instances compare to the benign instances. Previously, we showed this problem could be solved by creating several small sub-datasets and subsequently using those to train individual models, and finally reaching a result by performing a majority vote. Another way to deal with this problem is to generate synthetic data using a RBM and then using the synthetic data to balance the training dataset. 

In this section, we will discuss how synthetic data generated from a trained RBM has been used to balance the training dataset.
 A synthetic data sample can be generated from a RBM trained using CD-1 in the following way. We input a 64-bit vector formed using random 0s and 1s to a trained RBM. After 50 Gibbs cycles, we sample a 64-bit binary vector from the visible layer of the RBM. This sampled binary vector forms an instance of the synthetic dataset. Generating a synthetic dataset using QA is straightforward. One needs to embed a trained RBM onto the D-Wave quantum annealer and perform a quantum annealing step. For quantum annealing the anneal time was set to $20\mu s$ for each anneal and the number of samples that were requested was 10000.  Thus, 10000 samples can be obtained from the quantum annealer very quickly (1000 results within tens of milliseconds). From each sample, the states of the visible units are determined. Each sample corresponds to a record in the synthetic dataset. In this way, synthetic data composed of 10000 records is obtained using QA.

\begin{table*}[h!]
  \caption{\label{tab:gen_data}  Synthetic datasets are generated from RBMs trained using contrastive divergence (CD) and quantum annealing (QA). These synthetic datasets are then used to train RBMs using CD. The classification accuracies of these RBMs for benign and attack records, as well as total accuracies, are presented. These accuracies are calculated on the test dataset. The label `Model' indicates the RBM model that was used to generate the synthetic dataset.}
  \centering
    \begin{tabular}{|c|ccc|c|ccc|}
     \cline{1-8}
\noalign{\smallskip}
     \multirow{2}{*}{Model}    & \multicolumn{3}{c|}{Classification Accuracy } &  \multirow{2}{*}{Model} &
      \multicolumn{3}{c|}{Classification Accuracy } \\
\cline{2-4} \cline{6-8}
\noalign{\smallskip}
      &  Benign  (\%) &  Attack  (\%)& Total  (\%)  &   &Benign  (\%) &  Attack  (\%)& Total  (\%) \\
      \cline{1-8}
\noalign{\smallskip}
 A-CD & 78.99 & 80.91 & 79.31 & A-QA & 63.35  & 76.89 & 65.63 \\
 B-CD & 73.26 & 84.32 & 75.13 & B-QA & 67.34  & 73.62 & 68.40 \\
 C-CD & 59.35 & 90.66 & 64.63 & C-QA&72.28  & 66.73 & 79.42 \\
 D-CD & 61.69 & 87.12 & 65.98 & D-QA&72.89  & 69.05 & 72.24 \\
 E-CD & 68.13 & 85.28 & 71.02 & E-QA &61.51  &  78.19 & 64.32\\
\cline{1-8}
\noalign{\smallskip}
Average & 68.28 & 85.66 & 71.21 &Average& 67.47  & 72.90 & 70.00 \\
Stdev & 8.10 & 3.59 & 6.16 &Stdev & 5.12  & 4.93 & 6.08 \\
\cline{1-8}
      \end{tabular}
\end{table*}

\begin{figure*}
\begin{center}
\includegraphics[scale=1.0]{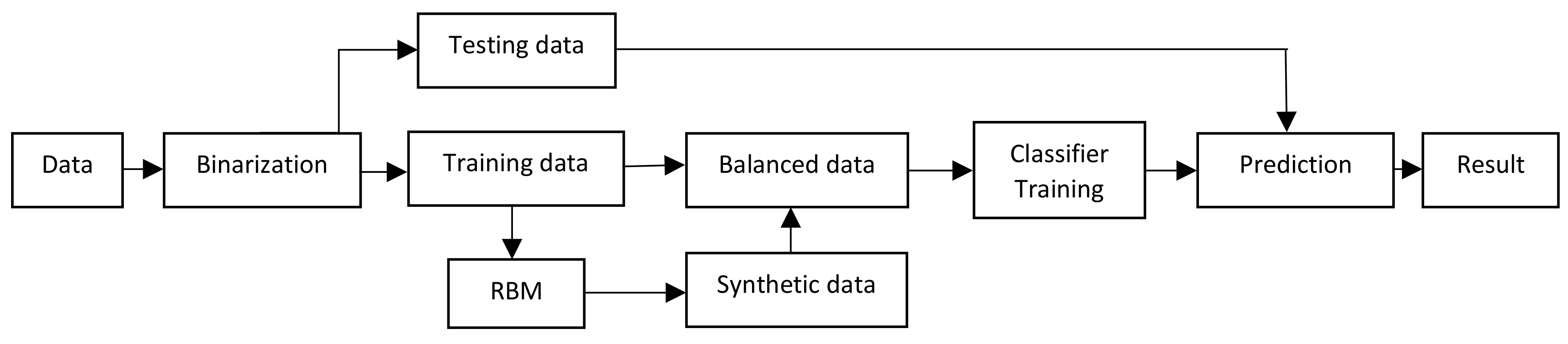}
\end{center}
\caption{\label{fig:S2} Scheme 2: Flowchart for intrusion detection using an imbalanced dataset. A RBM is first trained using the training dataset and then it is used to generate synthetic data. Training data and synthetic data are used to create a balanced dataset which is further used to train a classifier.} 
\end{figure*}

To ensure that the synthetic dataset generated from a trained RBM is useful, we perform  the following experiment. We use trained RBM models (A, B, C, D, and E) from the previous experiment to generate synthetic datasets; one from each model. Thus, ten datasets are generated; five from the CD-based RBMs and the other five from the QA-based RBMs. Now, ten RBM models are trained on these ten synthetic datasets using CD-1. The performance of these ten RBM models is compared by estimating classification accuracies on the test dataset composed of 8697 records (Benign=7230, Attack=1467).  The results from these RBM models as well as estimated average and standard deviation are presented in Table \ref{tab:gen_data}. RBMs trained using synthetic dataset generated from `CD-1 trained RBM' shows total classification accuracy varying between 64.63\% to 79.31\%. The  RBM trained with synthetic dataset obtained from `QA trained RBM' shows classification accuracy varying between 64.32\% to 79.42\%. The average classification accuracy for benign and attack classes are 68.28\% and 85.66\% with dataset obtained from CD-1 based RBM, and 67.47\% and 72.90\% with dataset obtained from QA based RBM. The results from Table \ref{tab:gen_data} indicate that useful synthetic data can be generated from a trained RBM. This synthetic data can be used to augment the original imbalanced training dataset in order to balance it. Also, on the basis of the classification accuracies, one can conclude that the samples obtained from a RBM trained using QA are as good as from a RBM trained with CD-1. 

Now we know that a RBM can be used to generate useful synthetic data. We can use this procedure to generate synthetic data to balance the training data and hence improve the performance of a classifier. Scheme 2, which uses a RBM to generate new data, is illustrated in Fig. \ref{fig:S2}. The original dataset is first binarized and divided into testing and training data. The training data is used to train a RBM, which is subsequently used to generate a synthetic dataset. Depending on the number of instances needed to balance the training data, one can use a subset of a synthetic dataset to balance the training dataset. There were 18000 benign and 3450 attack records in the training dataset, so 14550 synthetic attack records are added to balance the training dataset. A classifier is then trained on the balanced dataset and a prediction on the original testing data is performed.

\begin{table*}[h!]
  \caption{\label{tab:models}Balanced training data is used to train six classifiers. Performance metricses: precision, recall, $F_1$ score, and accuracy are used to compare models. The label `CD-bal' (`QA-bal') indicates that the synthetic data that is used to balance the training dataset is obtained from a RBM trained with contrastive divergence (quantum annealing). The label `imbal' indicates the original imbalanced dataset. Values are evaluated on the testing data.}
  \centering
 \begin{tabular}{|c|c|cc|cc|cc|c|}
     \hline
\noalign{\smallskip}
\multirow{2}{*}{Method}& \multirow{2}{*}{Data} &   \multicolumn{2}{c|}{Precision} & \multicolumn{2}{c|}{Recall} & \multicolumn{2}{c|}{$F_1$ score} &  \multirow{2}{*}{Accuracy, $A_{tot}$} \\
&  & Attack & Benign & Attack & Benign& Attack & Benign& \\
\hline
\noalign{\smallskip}

 Restricted  &  CD-bal &  0.87 & 0.94 & 0.95 & 0.85 & 0.91 & 0.89 & 85\% \\
Boltzmann &  QA-bal &  0.92 & 0.82 & 0.80 & 0.93 & 0.85 & 0.87 & 82\% \\
Machine&  imbal &  0.93 & 0.63 & 0.42 & 0.97 & 0.58 & 0.77 & 68\% \\
\hline
\noalign{\smallskip}
 Neural   & CD-bal  &  0.95 & 0.90 & 0.89 & 0.96 & 0.92 & 0.93 & 93\% \\
Network&  QA-bal &  0.95 & 0.90 & 0.89 & 0.96 & 0.92 & 0.93 & 93\% \\
(NN)&  imbal &  1.00 & 0.60 & 0.32 & 1.00 & 0.49 & 0.75 & 66\% \\
\hline
\noalign{\smallskip}
 K-Nearest  &  CD-bal  &  0.98 & 0.89 & 0.87 & 0.98 & 0.92 & 0.93 & 93\% \\
Neighbor&  QA-bal &  0.98 & 0.89 & 0.87 & 0.98 & 0.92 & 0.93 & 93\% \\
(KNN)&  imbal & 1.00  & 0.62 & 0.40 & 1.00 & 0.57 & 0.77 & 70\% \\
\hline
\noalign{\smallskip}
 Support&  CD-bal  &  0.94 & 0.77 & 0.70 & 0.95 & 0.80 & 0.86 & 83\% \\
Vector&  QA-bal &   0.94 & 0.77 & 0.70 & 0.95 & 0.80 & 0.85 & 83\% \\
Classifier&  imbal &  1.00 & 0.58 & 0.26 & 1.00 & 0.42 & 0.73 & 63\% \\
\hline
\noalign{\smallskip}
 \multirow{3}{*}{Decision Tree}   &  CD-bal  &  0.90 & 0.83 & 0.80 & 0.92 & 0.85 & 0.87 & 86\% \\
 &  QA-bal &  0.90 & 0.83 & 0.80 & 0.92 & 0.85 & 0.87 & 86\% \\
&  imbal &  1.00 & 0.56 & 0.22 & 1.00 & 0.36 & 0.72 & 61\% \\
\hline
\noalign{\smallskip}
 \multirow{3}{*}{Naive Bayes}   &  CD-bal  &  0.87 & 0.60 & 0.33 & 0.95 & 0.48 & 0.74 & 65\% \\
&  QA-bal &  0.87 & 0.60 & 0.33 & 0.95 & 0.48 & 0.74 & 65\% \\
&  imbal &  1.00 & 0.54 & 0.16 & 1.00 & 0.27 & 0.70 & 58\% \\
\hline

  \end{tabular}
\end{table*}

Considering the fact that RBMs are mostly used as a generative model and there are other classification methods that  perform better than a RBM classifier, we train several classifiers on the balanced training dataset. The results are presented  in Table \ref{tab:models}. For comparison, model performance with the original imbalanced dataset is also included. 
 We notice in the table that K Nearest Neighbor (KNN) and Neural Network (NN) performed better than other models. They both showed a classification accuracy of 93\%. Their values for precision, recall, and $F_1$ scores are also better than other methods. The lowest value of classification accuracy, as well as other metrics, are found in the case of Naive Bayes. The classification accuracy for this classifier is 65\%.  This exercise shows that it is important to investigate different classifiers to achieve better performance and different methods may give widely differing results. Table \ref{tab:models} shows that all classification methods show  improved performance when the dataset is balanced. Thus, the RBM-based technique that is used to balance the dataset using synthetic data is effective. This demonstrates the ability of the RBM to fill gaps in an imbalanced dataset by creating synthetic data that falls within the probability range of existing data.

\section{Discussion}

Quantum computers are still in a formative stage of their technology.  Consequently, comparing the RBM approach using QA to the mature classical CD or other approaches is uneven. The QA approach is expected to progress as quantum computing technology advances.  In scheme 1, we note that the total classification accuracies using QA-trained and CD-trained RBM are 80.04\% and 95.68\%, respectively. The performance gap that arises between CD-trained and QA-trained could be attributed to the following reasons. First, it has been observed by Koshka \emph{et~al.} \cite{koshka2020toward, k91} that RBM sampling using QA  misses many of the higher-energy regions of the configuration space, while also finding many new regions consistently missed by CD. Perhaps in the present case of the ISCX dataset, high energy samples missed by QA are also important. The overall effect is the RBM learns, but not as well as we expect. Another reason could be an instance-dependent effective temperature of the D-Wave annealer. We would like the D-Wave to sample with $kT=1$, where $T$ refers to the temperature at which the D-Wave operates. However, this is hardly the case and hence we introduce an effective scaling parameter $S$, for the Hamiltonian being embedded that allows us to ensure $SkT$ approximates unity. The effective scaling is treated as a hyperparameter and is fixed throughout the training of the RBM. Ideally one should calculate an effective temperature during each training epoch. This mismatch might degrade RBM's learning during the training. An accurate way to estimate the temperature at which the D-Wave samples for ground-state configuration is an open challenge. Efforts have been made towards identifying instance-dependent temperature for smaller models, none of which have proven to scale efficiently towards larger feature spaces \cite{PhysRevA.94.022308, caldeira2020}. Finally, hardware limitations like limited connectivity (which forces one to form long chains), quantum noise, low coherence time, etc could be some other reasons for the lower classification performance of the QA-based approach. 

When we compare synthetic data obtained from QA-trained RBM and CD-trained RBM (Table \ref{tab:gen_data}), we do not see much difference in classification performances. Accuracies of RBMs trained on both datasets are similar. These results indicate that our simplified approach of using a hyperparameter instead of an exact instance-dependent temperature is useful.
There is another advantage of using QA for RBM training. Depending on the complexity of a dataset, the CD might need hundreds of Gibbs cycles to reach the equilibrium to finally give one sample, while using a QA-based approach one can obtain 10000 samples almost instantaneously. Further, with the availability of quantum annealers with higher qubits and better connectivity, lower noise, the QA-based RBM training is likely to be improved and it would be possible to deal with larger datasets. Several investigators have shown that by employing machine learning techniques like principal component analysis and autoencoders to compress data, one can investigate a moderate size dataset with currently available quantum annealers \cite{caldeira2020, sleeman2020hybrid}. CD-based and QA-based approaches are fundamentally different ways of training RBMs. It would be an interesting exercise to train a RBM using samples obtained from both methods together. After training one should compare the  results with RBMs trained separately using QA and CD approaches. Our results indicate tha a RBM could be an effective tool to generate synthetic data that can be used to balance a dataset. One could also try training the RBM exclusively on the minority class to balance the original dataset. The QA-based approach can be used for faster sampling as sampling from a quantum annealer is almost instantaneous. 

We see that it is much easier to compute the model term using a D-Wave to sample low energy eigenstates of the Hamiltonians. This shows how a D-Wave machine can be utilized in problems beyond optimization and into the machine learning world. However, there are some limitations of  the quantum annealer that we  use.  The D-Wave 2000Q allows for a fully connected bilayer network of only 64 qubits at maximum. This limits the size of feature space of the data to be used for doing a study on large datasets without using another layer of feature extraction to downsize the data set used for the study. The qubits are noisy and less coherent compared to the upcoming new D-Wave machine which supports 5000 qubits and additional qubit interconnectivity and this provides opportunities for doing better analysis of the proposed schemes. A larger feature space would also allow for more confident claims to be made about the role that these machines might play in the machine learning world.

There are several advantages of using the D-Wave quantum annealer for RBM training. It offers a fundamentally different way to compute the model dependent term of the gradient of log-likelihood. Computation of this term using conventional methods like CD and PCD is intractable. Further, QA based sampling is faster than MCMC used in CD or PCD. So, we  expect that with improvement in hardware such as more qubits, lower noise, better coherence time  as well as robust algorithm for effective temperature, the QA-based RBM training is likely to perfom better than the CD-based approach.

The dataset that we use in this study is imbalanced. It comprises 30147 unique records. The number of records that belong to the attack class is 4917. It looks like the amount of data of the attack class  is not enough for RBM training. However, when we balance the dataset and train classifiers on it, the results indicate that the data amount is sufficient. For example, in the case of RBM trained on the “CD-bal” dataset, the precision, recall, and $F_1$ score for the attack class are 0.87, 0.95, and 0.91, respectively (Table \ref{tab:models}). It seems like though the number of attack records and features is small, the chosen records/features are representative of the model. Aldwairi \emph{et~al.} \cite{aldwairi2020m} established that when certain features that are representative of the model are to be selected, the change in the accuracy is minimal across all tested algorithms. Our first approach which uses under-sampling of benign records as well as the second approach where oversampling of attack records is used, seem to be effective for balancing the ISCX dataset.

\section{Conclusions}

Restricted Boltzmann machine (RBM) methodology  has been investigated for classification and synthetic data generation using the cybersecurity ISCX dataset. RBMs are trained through a quantum annealing approach performed using the D-Wave 2000Q quantum annealer.  For comparison, a state-of-the-art method for RBM training, contrastive divergence, is also investigated. The ISCX dataset is preprocessed and binarized to transform it into a form that can be used with a RBM. When a classifier is trained on the original data, it is found that attack records can be correctly predicted with an accuracy of 42\%, while benign records are predicted  with an accuracy of 97\%. This disproportionate result is attributed to the fact that the dataset is imbalanced. The attack records in the dataset only account for 14.1\% of the total number of records. To deal with the imbalanced dataset, we propose two schemes. The first scheme is based on the undersampling of benign  records. In this scheme, the training dataset  is divided into five sub-datasets. Five classifiers are trained separately on these datasets. The final result has been obtained by performing a majority voting on the results from the individual classifiers. Our results show that by using a majority vote the classification accuracy increased up to 95.68\% from 90.24\% in the case of CD-1. In the case of QA, the classification accuracy increased to 80.04\% from 74.14\%. The second scheme that we use to balance the training dataset  is based on the generation of synthetic data using a trained RBM.   The balanced dataset obtained from this scheme is used to train six different classifiers. Neural network and  K-nearest neighbor models perform the better than other classifiers. The results indicate that for the sampling applications, a RBM trained with QA is as good as a RBM trained with CD. Based on the classification accuracy results, we infer that both scheme 1 and scheme 2 significantly improved the classification accuracy compared to the case when the dataset was imbalanced.  The learning of QA-based RBM can be improved with the availability of improved quantum annealers with a large number of qubits as well as by using an efficient procedure to determine the  effective temperature of the QPU instead of treating it as a hyperparameter $S$.

\section*{Acknowledgment}

We are grateful for the support from Integrated Data Science Initiative Grants (IDSI F.90000303), Purdue University. S.K. would like to acknowledge funding by the U.S. Department of Energy (Office of Basic Energy Sciences) under Award No. DE-SC0019215.  T.S.H. would like to acknowledge funding by the U.S. Department of Energy, Office of Basic Energy Science. This research used resources of the Oak Ridge Leadership Computing Facility, which is a DOE Office of Science User Facility supported under Contract DE-AC05-00OR22725. We would like to thank  D-Wave RFP award under project `USRA NASA-AMES System Time'. This manuscript has been released as a pre-print at arXiv:2011.13996 [quant-ph]\cite{dixit2020b}. 

\section*{Author contributions}
 The research was planned by S.~Kais,  Y.~Koshka, M.A.~Novotny, T.S.~Humble, and M.A.~Alam. Algorithm development and calculations were performed by Vivek~Dixit and Raja~Selvarajan. Tamer~Aldwairi modified the original ISCX dataset into a 64-bit binary dataset. All authors contributed to the discussion of results and the writing of the manuscript.

\bibliographystyle{IEEEtran}

\begin{thebibliography}{10}
\providecommand{\url}[1]{#1}
\csname url@samestyle\endcsname
\providecommand{\newblock}{\relax}
\providecommand{\bibinfo}[2]{#2}
\providecommand{\BIBentrySTDinterwordspacing}{\spaceskip=0pt\relax}
\providecommand{\BIBentryALTinterwordstretchfactor}{4}
\providecommand{\BIBentryALTinterwordspacing}{\spaceskip=\fontdimen2\font plus
\BIBentryALTinterwordstretchfactor\fontdimen3\font minus
  \fontdimen4\font\relax}
\providecommand{\BIBforeignlanguage}[2]{{%
\expandafter\ifx\csname l@#1\endcsname\relax
\typeout{** WARNING: IEEEtran.bst: No hyphenation pattern has been}%
\typeout{** loaded for the language `#1'. Using the pattern for}%
\typeout{** the default language instead.}%
\else
\language=\csname l@#1\endcsname
\fi
#2}}
\providecommand{\BIBdecl}{\relax}
\BIBdecl

\bibitem{Smith-2020}
\BIBentryALTinterwordspacing
M.~Smith, Zhanna, and E.~Lostri, ``The hidden costs of cybercrime.''
  \emph{Technical Report. Santa Clara: McAfee.}, 2020. [Online]. Available:
  \url{https://www.mcafee.com/enterprise/en-us/assets/reports/rp-hidden-costs-of-cybercrime.pdf}
\BIBentrySTDinterwordspacing

\bibitem{fiore2013network}
U.~Fiore, F.~Palmieri, A.~Castiglione, and A.~De~Santis, ``Network anomaly
  detection with the restricted {B}oltzmann machine,'' \emph{Neurocomputing},
  vol. 122, pp. 13--23, 2013.

\bibitem{aldwairi2018evaluation}
T.~Aldwairi, D.~Perera, and M.~A. Novotny, ``An evaluation of the performance
  of restricted {B}oltzmann machines as a model for anomaly network intrusion
  detection,'' \emph{Computer Networks}, vol. 144, pp. 111--119, 2018.

\bibitem{alom2015intrusion}
M.~Z. Alom, V.~Bontupalli, and T.~M. Taha, ``Intrusion detection using deep
  belief networks,'' in \emph{2015 National Aerospace and Electronics
  Conference (NAECON)}.\hskip 1em plus 0.5em minus 0.4em\relax IEEE, 2015, pp.
  339--344.

\bibitem{salama2011hybrid}
M.~A. Salama, H.~F. Eid, R.~A. Ramadan, A.~Darwish, and A.~E. Hassanien,
  ``Hybrid intelligent intrusion detection scheme,'' in \emph{Soft computing in
  industrial applications}.\hskip 1em plus 0.5em minus 0.4em\relax Springer,
  2011, pp. 293--303.

\bibitem{li2015hybrid}
Y.~Li, R.~Ma, and R.~Jiao, ``A hybrid malicious code detection method based on
  deep learning,'' \emph{International Journal of Security and Its
  Applications}, vol.~9, no.~5, pp. 205--216, 2015.

\bibitem{alrawashdeh2016toward}
K.~Alrawashdeh and C.~Purdy, ``Toward an online anomaly intrusion detection
  system based on deep learning,'' in \emph{2016 15th IEEE international
  conference on machine learning and applications (ICMLA)}.\hskip 1em plus
  0.5em minus 0.4em\relax IEEE, 2016, pp. 195--200.

\bibitem{mizel2007simple}
A.~Mizel, D.~A. Lidar, and M.~Mitchell, ``Simple proof of equivalence between
  adiabatic quantum computation and the circuit model,'' \emph{Physical review
  letters}, vol.~99, no.~7, p. 070502, 2007.

\bibitem{farhi2000quantum}
E.~Farhi, J.~Goldstone, S.~Gutmann, and M.~Sipser, ``Quantum computation by
  adiabatic evolution,'' \emph{arXiv preprint quant-ph/0001106}, 2000.

\bibitem{aharonov:2007}
\BIBentryALTinterwordspacing
D.~Aharonov, W.~van Dam, J.~Kempe, Z.~Landau, S.~Lloyd, and O.~Regev,
  ``Adiabatic quantum computation is equivalent to standard quantum
  computation,'' \emph{SIAM Journal on Computing}, vol.~37, no.~1, pp.
  166--194, 2007. [Online]. Available:
  \url{https://doi.org/10.1137/S0097539705447323}
\BIBentrySTDinterwordspacing

\bibitem{Kadowaki:1998}
\BIBentryALTinterwordspacing
T.~Kadowaki and H.~Nishimori, ``Quantum annealing in the transverse {I}sing
  model,'' \emph{Phys. Rev. E}, vol.~58, pp. 5355--5363, Nov 1998. [Online].
  Available: \url{https://link.aps.org/doi/10.1103/PhysRevE.58.5355}
\BIBentrySTDinterwordspacing

\bibitem{FINNILA1994343}
\BIBentryALTinterwordspacing
A.~Finnila, M.~Gomez, C.~Sebenik, C.~Stenson, and J.~Doll, ``Quantum annealing:
  A new method for minimizing multidimensional functions,'' \emph{Chemical
  Physics Letters}, vol. 219, no.~5, pp. 343 -- 348, 1994. [Online]. Available:
  \url{http://www.sciencedirect.com/science/article/pii/0009261494001170}
\BIBentrySTDinterwordspacing

\bibitem{potok2020adiabatic}
T.~Potok \emph{et~al.}, ``Adiabatic quantum linear regression,'' \emph{arXiv
  preprint arXiv:2008.02355}, 2020.

\bibitem{Willsch2020}
\BIBentryALTinterwordspacing
D.~Willsch, M.~Willsch, H.~{De Raedt}, and K.~Michielsen, ``Support vector
  machines on the {D}-{W}ave quantum annealer,'' \emph{Computer Physics
  Communications}, vol. 248, p. 107006, 2020. [Online]. Available:
  \url{http://www.sciencedirect.com/science/article/pii/S001046551930342X}
\BIBentrySTDinterwordspacing

\bibitem{kumar}
V.~Kumar, G.~Bass, C.~Tomlin, and J.~Dulny, ``Quantum annealing for
  combinatorial clustering,'' \emph{Quantum Information Processing}, vol.~17,
  no.~2, p.~39, 2018.

\bibitem{das2019track}
S.~Das, A.~J. Wildridge, S.~B. Vaidya, and A.~Jung, ``Track clustering with a
  quantum annealer for primary vertex reconstruction at hadron colliders,''
  \emph{arXiv preprint arXiv:1903.08879}, 2019.

\bibitem{arthur2020}
D.~Arthur \emph{et~al.}, ``Balanced k-means clustering on an adiabatic quantum
  computer,'' \emph{arXiv preprint arXiv:2008.04419}, 2020.

\bibitem{Kais2018a}
S.~Jiang, K.~A. Britt, A.~J. McCaskey, T.~S. Humble, and S.~Kais, ``Quantum
  annealing for prime factorization,'' \emph{Nature Scientific reports}, 2018.

\bibitem{Kais:2018}
\BIBentryALTinterwordspacing
R.~Xia, T.~Bian, and S.~Kais, ``Electronic structure calculations and the
  {I}sing hamiltonian,'' \emph{The Journal of Physical Chemistry B}, vol. 122,
  no.~13, pp. 3384--3395, 2018. [Online]. Available:
  \url{https://doi.org/10.1021/acs.jpcb.7b10371}
\BIBentrySTDinterwordspacing

\bibitem{adachi2015application}
S.~H. Adachi and M.~P. Henderson, ``Application of quantum annealing to
  training of deep neural networks,'' \emph{arXiv preprint arXiv:1510.06356},
  2015.

\bibitem{PhysRevA.94.022308}
\BIBentryALTinterwordspacing
M.~Benedetti, J.~Realpe-G\'omez, R.~Biswas, and A.~Perdomo-Ortiz, ``Estimation
  of effective temperatures in quantum annealers for sampling applications: A
  case study with possible applications in deep learning,'' \emph{Phys. Rev.
  A}, vol.~94, p. 022308, Aug 2016. [Online]. Available:
  \url{https://link.aps.org/doi/10.1103/PhysRevA.94.022308}
\BIBentrySTDinterwordspacing

\bibitem{koshka2020toward}
Y.~Koshka and M.~A. Novotny, ``Toward sampling from undirected probabilistic
  graphical models using a {D}-{W}ave quantum annealer,'' \emph{Quantum
  Information Processing}, vol.~19, no.~10, pp. 1--23, 2020.

\bibitem{k91}
Y.~{Koshka} and M.~A. {Novotny}, ``Comparison of {D}-{W}ave quantum annealing
  and classical simulated annealing for local minima determination,''
  \emph{IEEE Journal on Selected Areas in Information Theory}, vol.~1, no.~2,
  pp. 515--525, 2020.

\bibitem{dixit2020training}
V.~Dixit, R.~Selvarajan, M.~A. Alam, T.~S. Humble, and S.~Kais, ``Training and
  classification using a restricted {B}oltzmann machine on the {D}-{W}ave
  {2000Q},'' \emph{arXiv preprint arXiv:2005.03247}, 2020.

\bibitem{caldeira2020}
J.~Caldeira, J.~Job, S.~H. Adachi, B.~Nord, and G.~N. Perdue, ``Restricted
  {B}oltzmann machines for galaxy morphology classification with a quantum
  annealer,'' \emph{arXiv preprint arXiv:1911.06259}, 2019.

\bibitem{sleeman2020hybrid}
J.~Sleeman, J.~Dorband, and M.~Halem, ``A hybrid quantum enabled {RBM}
  advantage: convolutional autoencoders for quantum image compression and
  generative learning,'' in \emph{Quantum Information Science, Sensing, and
  Computation XII}, vol. 11391.\hskip 1em plus 0.5em minus 0.4em\relax
  International Society for Optics and Photonics, 2020, p. 113910B.

\bibitem{hinton:2002}
\BIBentryALTinterwordspacing
G.~E. Hinton, ``Training products of experts by minimizing contrastive
  divergence,'' \emph{Neural Computation}, vol.~14, no.~8, pp. 1771--1800,
  2002. [Online]. Available: \url{https://doi.org/10.1162/089976602760128018}
\BIBentrySTDinterwordspacing

\bibitem{chawla2002smote}
N.~V. Chawla, K.~W. Bowyer, L.~O. Hall, and W.~P. Kegelmeyer, ``Smote:
  synthetic minority over-sampling technique,'' \emph{Journal of artificial
  intelligence research}, vol.~16, pp. 321--357, 2002.

\bibitem{fernandez2018smote}
A.~Fern{\'a}ndez, S.~Garcia, F.~Herrera, and N.~V. Chawla, ``Smote for learning
  from imbalanced data: progress and challenges, marking the 15-year
  anniversary,'' \emph{Journal of artificial intelligence research}, vol.~61,
  pp. 863--905, 2018.

\bibitem{ma2020aesmote}
X.~Ma and W.~Shi, ``Aesmote: Adversarial reinforcement learning with smote for
  anomaly detection,'' \emph{IEEE Transactions on Network Science and
  Engineering}, 2020.

\bibitem{yan2017novel}
B.~Yan, G.~Han, M.~Sun, and S.~Ye, ``A novel region adaptive smote algorithm
  for intrusion detection on imbalanced problem,'' in \emph{2017 3rd IEEE
  International Conference on Computer and Communications (ICCC)}.\hskip 1em
  plus 0.5em minus 0.4em\relax IEEE, 2017, pp. 1281--1286.

\bibitem{su2018research}
P.~Su, Y.~Liu, and X.~Song, ``Research on intrusion detection method based on
  improved smote and {XGBoost},'' in \emph{Proceedings of the 8th International
  Conference on Communication and Network Security}, 2018, pp. 37--41.

\bibitem{tesfahun2013intrusion}
A.~Tesfahun and D.~L. Bhaskari, ``Intrusion detection using random forests
  classifier with smote and feature reduction,'' in \emph{2013 International
  Conference on Cloud \& Ubiquitous Computing \& Emerging Technologies}.\hskip
  1em plus 0.5em minus 0.4em\relax IEEE, 2013, pp. 127--132.

\bibitem{ahsan2018smote}
M.~Ahsan, R.~Gomes, and A.~Denton, ``Smote implementation on phishing data to
  enhance cybersecurity,'' in \emph{2018 IEEE International Conference on
  Electro/Information Technology (EIT)}.\hskip 1em plus 0.5em minus 0.4em\relax
  IEEE, 2018, pp. 0531--0536.

\bibitem{shiravi2012toward}
A.~Shiravi, H.~Shiravi, M.~Tavallaee, and A.~A. Ghorbani, ``Toward developing a
  systematic approach to generate benchmark datasets for intrusion detection,''
  \emph{Computers \& Security}, vol.~31, no.~3, pp. 357--374, 2012.

\bibitem{sutskever2010}
I.~Sutskever and T.~Tieleman, ``On the convergence properties of contrastive
  divergence,'' in \emph{Proceedings of the thirteenth international conference
  on artificial intelligence and statistics}, 2010, pp. 789--795.

\bibitem{Dumoulin}
\BIBentryALTinterwordspacing
V.~Dumoulin, A.~C. Ian J~Goodfellow, and Y.~Bengio, ``On the challenges of
  physical implementations of {RBM}s,'' \emph{AAAI Publications, Twenty-Eighth
  AAAI Conference on Artificial Intelligence}, 2014. [Online]. Available:
  \url{https://www.aaai.org/ocs/index.php/AAAI/AAAI14/paper/viewPaper/8608}
\BIBentrySTDinterwordspacing

\bibitem{Humble:2018}
\BIBentryALTinterwordspacing
T.~D. Goodrich, B.~D. Sullivan, and T.~S. Humble, ``Optimizing adiabatic
  quantum program compilation using a graph-theoretic framework,''
  \emph{Quantum Information Processing}, vol.~17, April 2018. [Online].
  Available: \url{https://doi.org/10.1007/s11128-018-1863-4}
\BIBentrySTDinterwordspacing

\bibitem{koshka2017determination}
Y.~Koshka, D.~Perera, S.~Hall, and M.~A. Novotny, ``Determination of the
  lowest-energy states for the model distribution of trained restricted
  {B}oltzmann machines using a 1000 qubit {D}-{W}ave 2x quantum computer,''
  \emph{Neural Computation}, vol.~29, no.~7, pp. 1815--1837, 2017.

\bibitem{koshka2018comparison}
Y.~Koshka and M.~A. Novotny, ``Comparison of use of a 2000 qubit {D}-{W}ave
  quantum annealer and {MCMC} for sampling, image reconstruction, and
  classification,'' \emph{IEEE Transactions on Emerging Topics in Computational
  Intelligence}, vol.~5, no.~1, pp. 119--129, 2021.

\bibitem{scikit-learn}
F.~Pedregosa, G.~Varoquaux, A.~Gramfort, V.~Michel, B.~Thirion, O.~Grisel,
  M.~Blondel, P.~Prettenhofer, R.~Weiss, V.~Dubourg, J.~Vanderplas, A.~Passos,
  D.~Cournapeau, M.~Brucher, M.~Perrot, and E.~Duchesnay, ``Scikit-learn:
  Machine learning in {P}ython,'' \emph{Journal of Machine Learning Research},
  vol.~12, pp. 2825--2830, 2011.

\bibitem{witten2002data}
I.~H. Witten and E.~Frank, ``Data mining: practical machine learning tools and
  techniques with java implementations,'' \emph{Acm Sigmod Record}, vol.~31,
  no.~1, pp. 76--77, 2002.

\bibitem{aldwairi2020m}
T.~Aldwairi, D.~Perera, and M.~A. Novotny, ``Measuring the impact of accurate
  feature selection on the performance of {RBM} in comparison to state of the
  art machine learning algorithms,'' \emph{Electronics}, vol.~9, no.~7, p.
  1167, 2020.

\bibitem{dixit2020b}
V.~Dixit, R.~Selvarajan, M.~A. Alam, T.~S. Humble, and S.~Kais, ``Training a
  quantum annealing based restricted {B}oltzmann machine on cybersecurity
  data,'' \emph{arXiv preprint arXiv:2011.13996}, 2020.

\end{thebibliography}



\begin{IEEEbiography}
    [{\includegraphics[width=1in,height=1.25in]{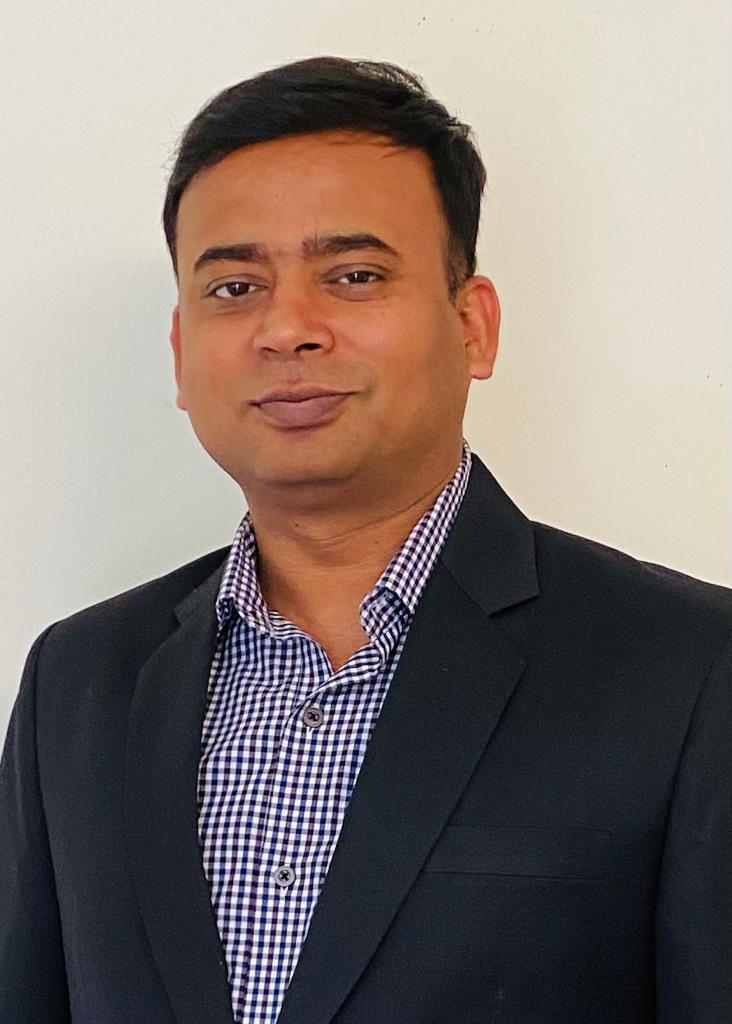}}]{Vivek Dixit}
received the M.S. and  Ph.D. degrees in physics and applied physics both from Mississippi State University, Starkville, MS, USA in 2011, and 2015, respectively. He is a postdoctoral research associate in the department of chemistry, Purdue University, West Lafayette, IN, USA. His research interests include Quantum Machine Learning, Quantum Computing, electronic structure calculations of solids and molecular systems, and spectroscopy.
\end{IEEEbiography}
\vspace{-15 mm}
\begin{IEEEbiography}
    [{\includegraphics[width=1in,height=1.25in]{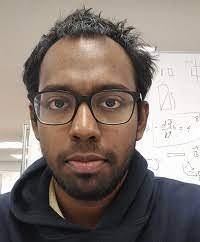}}]{Raja Selvarajan}
is a graduate student in the department of physics and astronomy, Purdue University, West Lafayette, IN, USA. His research interests include Quantum Machine Learning and Quantum Computing.
\end{IEEEbiography}
\vspace{-15 mm}
\begin{IEEEbiography}
    [{\includegraphics[width=1in,height=1.25in]{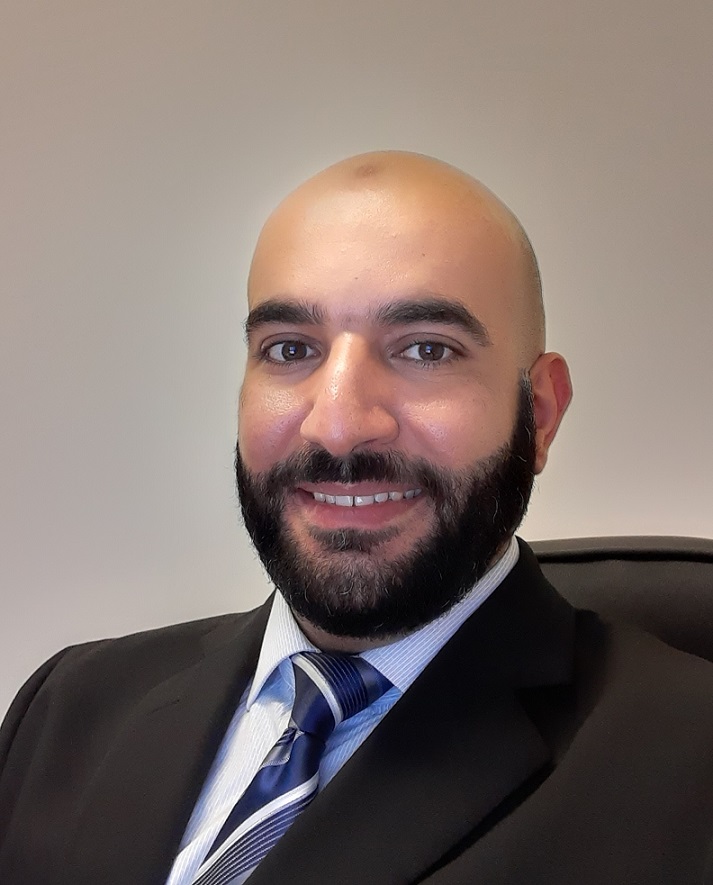}}]{Tamer Aldwairi}
received his M.S. in computer science and Ph.D. in computational engineering from Mississippi State University, MS, USA, in 2014. He worked as a Postdoc research associate at the Distributed Analytics and Security Institute in the High-Performance Computing Collaboratory (HPCC) from 2015 - 2017. He worked as a visiting assistant professor at Ursinus College from 2017 - 2019. He is currently an assistant professor of Instruction at Temple University. His research interests include machine learning, big data analytics, cybersecurity, quantum computing, and high-performance computing. 
\end{IEEEbiography}
\vspace{-15 mm}
\begin{IEEEbiography}
    [{\includegraphics[width=1in,height=1.25in]{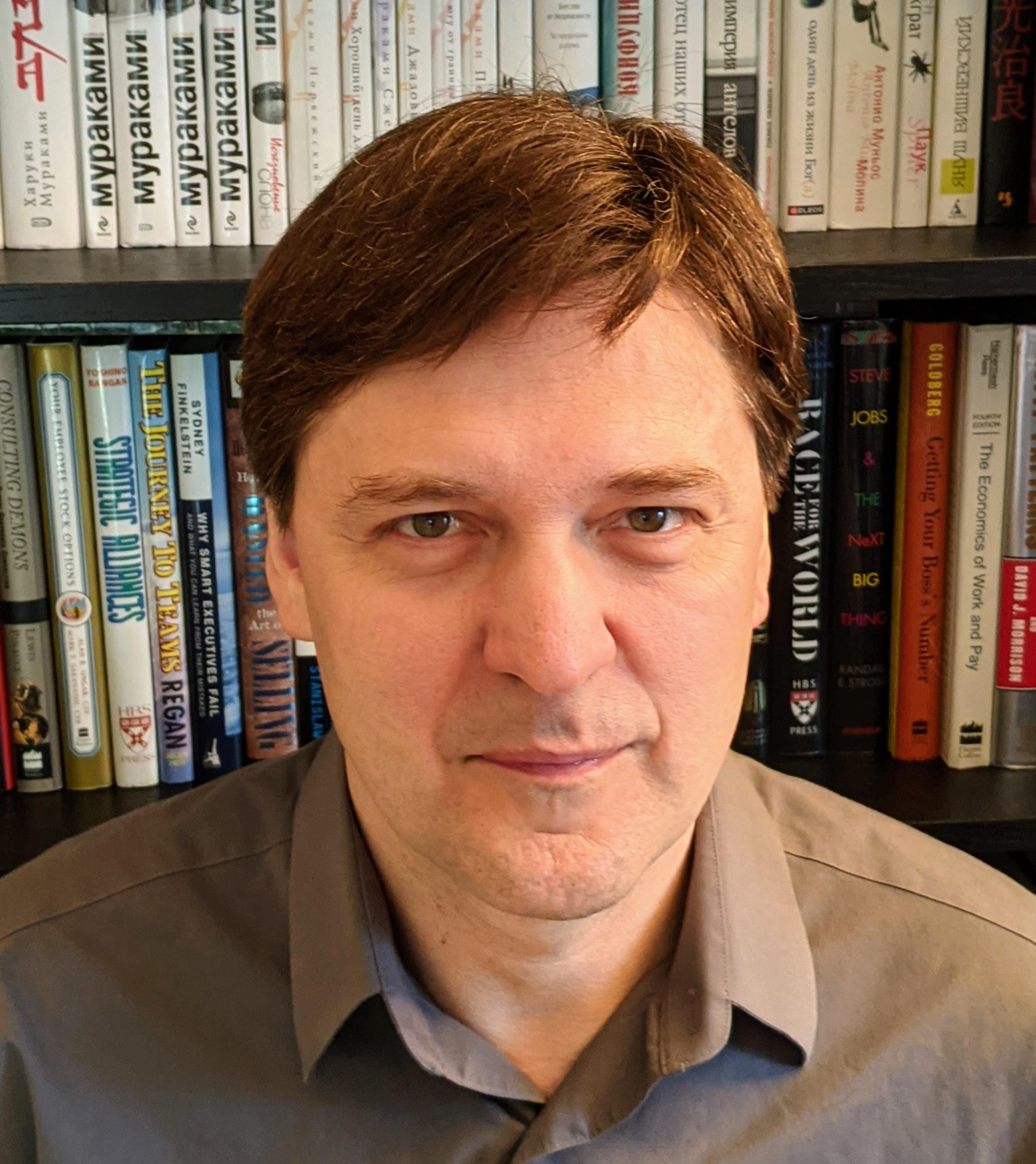}}]{Yaroslav Koshka}
 received the B.S. and M.S. degrees in electronics from Kiev
Polytechnic University, Kiev, Ukraine, in 1993, and the Ph.D. degree in electrical
engineering from the University of South Florida, Tampa, FL, USA, in 1998.
He is currently a Professor with the Department of Electrical and Computer Engineering, MSU, Starkville, MS,
USA and the Director of the Emerging Materials Research Laboratory. His current
main research area is quantum computations, their application to machine
learning as well as to properties of electronic materials. Other research interests
include semiconductor materials and device characterization, defect engineering,
synthesis of wide-bandgap semiconductor materials and nanostructures,
physics of semiconductor devices, and nanoelectronics.
\end{IEEEbiography}
\vspace{-15 mm}
\begin{IEEEbiography}
    [{\includegraphics[width=1in,height=1.25in]{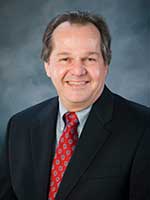}}]{Mark A. Novotny}
 received the B.S. degree in physics from North Dakota State
University, Fargo, ND, USA, in 1973, and the Ph.D. degree in physics from
Leland Stanford Junior University, Stanford, CA, USA, in 1979.
Since 2001, he has been a Professor and the Head of the Department of
Physics and Astronomy, Mississippi State University, Starkville, MS, USA. His
research interests include classical and quantum computational physics, with
applications to classical and quantum properties of materials and statistical
mechanics. He has authored or coauthored more than 200 papers in refereed
journals. Dr. Novotny has the honor of being a Giles Distinguished Professor at
Mississippi State University, as well as being a Fellow of the American Physical
Society and a Fellow of AAAS.
\end{IEEEbiography}
\vspace{-5 mm}
\begin{IEEEbiography}
    [{\includegraphics[width=1in,height=1.25in]{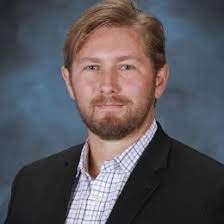}}]{Travis S. Humble} 
 is Deputy Director at the Department of Energy’s Quantum Science Center, a Distinguished Scientist at Oak Ridge National Laboratory, and director of the lab’s Quantum Computing Institute. Travis is leading the development of new quantum technologies and infrastructure to impact the DOE mission of scientific discovery through quantum computing. He is editor-in-chief for ACM Transactions on Quantum Computing, Associate Editor for Quantum Information Processing, and co-chair of the IEEE Quantum Initiative. Travis also holds a joint faculty appointment with the University of Tennessee Bredesen Center for Interdisciplinary Research and Graduate Education, where he works with students in developing energy-efficient computing solutions. He received his doctorate in theoretical chemistry from the University of Oregon before joining ORNL in 2005. 
\end{IEEEbiography}
\vspace{-15 mm}
\begin{IEEEbiography}
    [{\includegraphics[width=1in,height=1.25in]{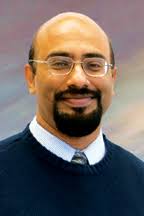}}]{Muhammad A. Alam} 
(M96, SM01, F06) is the Jai N. Gupta professor of Electrical Engineering at Purdue University, where his research focuses on physics, simulation, characterization and technology of classical and emerging electronic devices, including reliability of scaled MOSFET, theoretical foundation of nano-bio sensors, and atom-to-farm performance and modeling of solar cells. His work has been recognized by 2006 IEEE Kiyo Tomiyasu Medal, 2015 SRC Technical Excellence Award, and 2018 IEEE EDS Education Award.
\end{IEEEbiography}
\vspace{-15 mm}
\begin{IEEEbiography}
    [{\includegraphics[width=1in,height=1.25in]{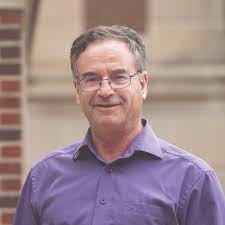}}]{Sabre Kais}  received the BSc, MSc, and PhD degrees at the Hebrew University of Jerusalem in 1983, 1984, and 1989 respectively. From 1989 to 1994, he was a research associate at Harvard University, Department of Chemistry. He joined Purdue University in1994 as an assistant professor of theoretical chemistry. Currently, he is a full professor of chemical physics, a professor of computer science (courtesy), a professor of physics at Purdue University. He has published over 240 papers in peer-reviewed journals. His research interests mainly include finite size scaling, dimensional scaling, Quantum Information, and Quantum Computing, He was the director of NSF funded center of innovation on “Quantum Information for Quantum Chemistry”, from 2010-2013. He is a Fellow of the American Physical Society, Fellow of the American Association for the Advancement of Science, Guggenheim Fellow, Purdue University Faculty Scholar, National Science Foundation Career Award Fellow He is a recipient of the 2012 Sigma Xi Research Award, and 2019 Herbert Newby McCoy Award, Purdue University.
\end{IEEEbiography}






\end{document}